\documentclass[aps,prb,reprint,superscriptaddress,footnotebib,shortbibliography]{revtex4-2}

\usepackage{amssymb}
\usepackage{amsmath,bm}
\usepackage{braket}
\usepackage{graphicx}

\usepackage[dvipsnames]{xcolor}

\usepackage{hyperref}
\hypersetup{%
	pdftoolbar=true,        
	pdfmenubar=true,        
	pdffitwindow=false,     
	pdfstartview={FitH},    
	pdftitle={title},    
	pdfauthor={Gerard Valent\'i-Rojas},     
	pdfsubject={Subject},   
	pdfcreator={Gerard Valent\'i-Rojas},   
	pdfproducer={Gerard Valent\'i-Rojas}, 
	pdfnewwindow=true,      
	colorlinks=true,       
	linkcolor=RoyalBlue,          
	citecolor=OrangeRed,        
	filecolor=Green,      
	urlcolor=RoyalBlue           
}

\usepackage{lipsum}

\usepackage{newtxtext,newtxmath}

\begin{document}


\title{Synthetic Flux Attachment}


\author{Gerard Valent\'i-Rojas}
\email[Corresponding author: ]{gv16@hw.ac.uk}
\affiliation{SUPA, Institute of Photonics and Quantum Sciences, Heriot-Watt University, Edinburgh, EH14 4AS, United Kingdom}

\author{Niclas Westerberg}
\affiliation{School of Physics \& Astronomy, University of Glasgow, Glasgow G12 8QQ, UK}
\affiliation{SUPA, Institute of Photonics and Quantum Sciences, Heriot-Watt University, Edinburgh, EH14 4AS, United Kingdom}

\author{Patrik \"Ohberg}
\affiliation{SUPA, Institute of Photonics and Quantum Sciences, Heriot-Watt University, Edinburgh, EH14 4AS, United Kingdom}


\date{\today}

\begin{abstract}
Topological field theories emerge at low energy in strongly-correlated condensed matter systems and appear in the context of planar gravity. In particular, the study of Chern-Simons terms gives rise to the concept of flux attachment when the gauge field is coupled to matter, yielding flux-charge composites. We investigate the generation of flux attachment in a Bose-Einstein condensate in the presence of non-linear synthetic gauge potentials. In doing so, we identify the $U(1)$ Chern-Simons gauge field as a singular density-dependent gauge potential, which in turn can be expressed as a Berry connection. We envisage a proof-of-concept scheme where the artificial gauge field is perturbatively induced by an effective light-matter detuning created by interparticle interactions. At a mean field level, we recover the action of a "charged" superfluid minimally coupled to both a background and a Chern-Simons gauge field. Remarkably, a localised density perturbation in combination with a non-linear gauge potential gives rise to an effective composite boson model of fractional quantum Hall effect, displaying anyonic vortices.

\end{abstract}


\maketitle

\section{Introduction}

Gauge invariance constitutes a conceptual cornerstone in the modern description of fundamental interactions of Nature \cite{yang2014conceptual, gross1992gauge, gross1996role, jackson2001historical, o2000gauge}. The mathematical structure obeying the principle that physics must not change from point to point in space and time hides a redundancy. This translates into a descriptive freedom of choice that must not affect the real world. Thus, only objects that are invariant under a gauge transformation are physical. However, this does not imply that gauge-dependent quantities are irrelevant. This statement is beautifully illustrated in quantum mechanics by the Aharonov-Bohm effect \cite{aharonov1959significance}. The wavefunction of a charged particle moving in a region of non-vanishing vector potential, but in which the magnetic field is zero everywhere except for at a single point, may pick up a global phase factor, yielding measurable phase shifts in an interference experiment \cite{chambers1960shift, fowler1961electron, boersch1961experimenteller, tonomura1982observation, tonomura1986evidence, osakabe1986experimental}. This has been instrumental in adopting the concepts of \textit{fibre bundles} and \textit{connections} \cite{wu1975concept} in the physics community. The Aharonov-Bohm phase constitutes an example of the more general concept of \textit{geometric (Berry) phase} \cite{pancharatnam1956generalized,mead1979determination,berry1984quantal, cohen2019geometric},  which has been particularly useful in the understanding of topological phases of matter, an intensively studied field in the last decade \cite{hasan2010colloquium, qi2011topological, senthil2015symmetry, ando2015topological, chiu2016classification, wen2017colloquium, armitage2018weyl, zhang2018topological, ozawa2019topological}. 

The recent ability to engineer \textit{artificial} (also known as \textit{synthetic}) gauge potentials in a variety of setups \cite{aidelsburger2018artificial} has made it possible to extend this exploration both to classical \cite{huber2016topological} and quantum simulators \cite{georgescu2014quantum}. In particular, the exquisite control and flexibility offered by ultracold atoms, and the possibility of tuning interactions, makes them an ideal setup for mimicking intriguing phenomena \cite{bloch2008many}. Creating gauge potentials in both optical lattices and the continuum is currently possible in multiple ways \cite{dalibard2011colloquium,dd2015introduction} by, for instance, rotation of the atomic gas \cite{fetter2009rotating}, using time-periodic drivings \cite{goldman2014periodically}, or light-induced methods \cite{goldman2014light}. Notwithstanding, these are in general non-dynamical or background gauge fields. This implies that the gauge fields do not have an equation of motion, and thus, are not representing gauge theories, but models with matter coupled to gauge potentials. This is an obstacle if one aims to emulate scenarios that require back-reaction of matter onto the gauge field, such as the quantum simulation of gauge theories or dynamical curved spacetimes.


Thus, great effort is currently being put in giving dynamics to synthetic gauge potentials \cite{banuls2019review,banuls2019simulating}. The usual top-down approach \cite{wiese2013ultracold,zohar2015quantum,dalmonte2016lattice}, using the Kogut-Susskind formalism \cite{kogut1975hamiltonian} and \textit{quantum link} models \cite{chandrasekharan1997quantum}, builds on the knowledge gathered from lattice gauge theories \cite{kogut1979introduction}. This requires some approximations, such as truncation of the Hilbert space, but has allowed for the first digital experimental realisation of the Schwinger model in (1+1)D \cite{martinez2016real} with trapped ions. Minimal building blocks for an analogue simulation of the same model in atomic mixtures have also been reported recently \cite{mil2020scalable}, as well as large-scale manufacturing of local constraints for bosons in optical superlattices \cite{yang2020observation}. Higher dimensional models, however, are still awaiting a realisation, mainly because of experimental challenges in controlling plaquette terms and the implementation of local constraints. 


On the other hand, a bottom-up approach starting from background gauge fields could also be possible \footnote{Recent alternative proposals are already making progress. In \cite{celi2019emerging}, the authors consider a system of Rydberg atoms in configurable arrays where, by exploiting a duality mapping of the Rokhsar-Kivelson model on the square lattice --- \textit{i.e.} a $U(1)$ lattice gauge theory in (2+1)D ---, they are able to reformulate the plaquette interactions as Rabi oscillations subject to Rydberg  blockade, which is desirable in view of potential experimental realisations.}. The main challenges are identifying and incorporating the minimal ingredients for a gauge theory in the formalism. First efforts for delivering back-action between the matter and gauge sectors in this sense are the so-called \textit{density-dependent gauge potentials} \cite{keilmann2011statistically,edmonds2013simulating,greschner2014density}. Only very recently, the first lattice gauge theory \cite{barbiero2019coupling} within this approach has been put forward, relying on inter-species density-dependent Peierls phases as carriers of the gauge interaction. Recent experimental studies show that such gauge potentials are within reach \cite{clark2018observation,gorg2019realization,schweizer2019floquet,lienhard2020realization}. Despite this, an issue not yet resolved is how do these density-dependent gauge potentials fit in this classification, provided that they are neither background fields nor do they yield a complete gauge theory \textit{per se}. More importantly, what are they useful for, and are there any physical system where these are present? So far, density-dependent gauge fields have been used in the context of pseudo-linear "anyons" \cite{keilmann2011statistically, greschner2015anyon, strater2016floquet}, as a mechanism to induce frustration on a lattice \cite{mishra2016density}, and in condensates with exotic phenomenology \cite{zheng2015topological, edmonds2015elementary,butera2016vortex,ohberg2019quantum, bermudez2015interaction, dingwall2019stability, butera2018curved, edmonds2020vortex}. \\

In this work we study the connection between density-dependent gauge potentials and topological field theories. In particular, we show that the Abelian Chern-Simons gauge field can be reinterpreted as a singular density-dependent gauge potential. From this, it follows that a $U(1)$ Chern-Simons term can be engineered by means of synthetic gauge fields with a nonlocal vortex-like kernel. We argue that this is ensured by the so-called \textit{flux attachment} constraint. Furthermore, we illustrate this idea by means of a proof-of-principle calculation for an experimentally feasible scheme to generate flux attachment. Starting from a microscopic Hamiltonian, we derive a mean-field theory for a Bose-Einstein condensate minimally coupled to a density-dependent Berry connection. As we will see, the latter plays the role of a synthetic gauge potential. We find that fine-tuning of the laser parameters allows for flux attachment without the need of long-range interactions. We recover an emergent effective description in the form of a Chern-Simons coupled superfluid action.

The relevance of our findings is two-fold. On the one hand,  at a practical level we theoretically describe a way to microscopically engineer a term that is typically \textit{emergent}, meaning that it appears effectively as a consequence of the collective rearrangement of a quantum many-body system. On the other hand, at a conceptual level we identify Chern-Simons as a theory involving density-dependent gauge fields. This is connected to well-known examples of systems that harbour such gauge fields \cite{wen1990topological,sachdev2018topological}, namely \textit{topologically ordered} (TO) matter \footnote{As opposed to \textit{symmetry protected topological} (SPT) phases of matter, understood as a minimal generalisation of the notion of a topological insulator. We consider the notion of topological order in Wen's modern classification of gapped topological phases.}. Thus, this helps to bridge the gap between background gauge fields and gauge theories in the context of quantum simulation, and at the same time explains why we should expect that density-dependent gauge fields come hand in hand with the appearance of anyons.

The outline of the paper is as follows. For the purpose of being self-contained and self-consistent, in section \ref{sec:cs_section} we review the importance and the main features of the Abelian Chern-Simons theory, and define the notion of flux attachment for this work. In section \ref{sec:flux_sense} we reinterpret the concept of flux attachment in the context of geometric phases in the so-called flux-tube or composite particle picture. This view is naturally related to artificial gauge fields in section \ref{sec:to_qsim}, where we discuss a possible experimental realisation. We proceed in section \ref{sec:dens_dep_bec} to introduce our model for a flux-attached bosonic field as a Bose-Einstein condensate subject to an effective Berry connection which depends on interparticle interactions. Then, in section \ref{sec:consequences}, we briefly analyse the direct implications of our results. Finally, in section \ref{sec:conclusion} we summarise our findings and discuss their implications.


\section{Revisiting the Abelian Chern-Simons Term}\label{sec:cs_section}

Low-dimensional physics has sparked an increasing amount of interest in the recent years in diverse contexts mainly due to the integrability of some models, and unusual phenomena sensitive to dimensionality. The latter is related to topological systems, and a primary example is Chern-Simons theory \cite{dunne1999aspects}, which has been the subject of study for the last 40 years. It has been used as a mechanism to make gauge fields massive \cite{schonfeld1981mass,deser1982topomass}, as a modification of General Relativity \cite{deser1982three, jackiw2003chern, alexander2009chern}, as an exactly solvable toy model for quantum gravity  \cite{witten1988}, as a way to generate self-dual vortices \cite{jackiw1990classical,jackiw1990soliton,jackiw1992self}, or as a low-energy effective theory of the fractional quantum Hall effect (QHE) \cite{girvin1987off, zhang1989effective,read1989order,lopez1991fractional}. More recently, there has been a revival in the more general context of topologically ordered states \cite{wen1990topological,wen1991topological,hansson2004superconductors,xu2009global,sachdev2018topological}, fractional topological and Chern insulators \cite{neupert2015fractional}, and the theory of composite Fermi liquids \cite{halperin1993theory,son2015composite,barkeshli2015particle}. This has, in turn, inspired a new family of particle-vortex dualities \cite{metlitski2016particle,wang2015dual,mross2016explicit}, which have been shown to fit in an even larger web of dual models  \cite{karch2016particle,seiberg2016duality,murugan2017particle}. These works provide a modern and unified view of the phenomenology of Chern-Simons theory as a multifaceted construction, encapsulating the pathway between a "particle" face of the duality and a "vortex" counterpart via the mechanism of flux attachment.

\subsection{Pure Abelian Chern-Simons}
We consider a $U(1)$ gauge field $A^{\,\mu} =(A^{\,0},\,\mathbf{A})$ in 2+1 dimensional spacetime ($\mu = 0,1,2$). We will use $c = 1$ unless explicitly noted otherwise, greek indices for spacetime components, and latin indices for space-only components. The Abelian Chern-Simons action is given by
\begin{equation}\label{eq:cs_lag}
S_{\mathrm{CS}} = \int dt\,d^{2}\mathbf{r}\; \mathcal{L}_{\mathrm{CS}} = \frac{\kappa e^{2}}{4 \pi \hbar} \int dt\,d^{2}\mathbf{r}\; \epsilon^{\,\mu \nu \lambda} A_{\,\mu}\, \partial_{\nu}\,A_{\lambda} \;,
\end{equation}
where $\kappa$ is a dimensionless coefficient often called the Chern-Simons level, and $\epsilon^{\,\mu \nu \lambda} $ is the Levi-Civita symbol. The Lagrangian density $\mathcal{L}_{CS}$ in \eqref{eq:cs_lag} is local, Lorentz invariant, and $\mathcal{PT}$ symmetric (although breaks separately $\mathcal{P}$ and $\mathcal{T}$). After a gauge transformation of the form $A_{\mu} \rightarrow A_{\mu} + \partial_{\mu}\, \Lambda $, it yields boundary terms like
\begin{equation}
\delta \mathcal{L}_{\mathrm{CS}} = \frac{\kappa e^{2}}{4 \pi \hbar}\, \partial_{\mu} \,(\Lambda\, \epsilon^{\,\mu \nu \lambda} \partial_{\nu}\,A_{\lambda}) \,.
\end{equation}
If boundaries can be neglected \footnote{This may not be the case in some situations, including finite samples, for which a bulk-edge correspondence is found. However, we will not consider those scenarios here.}, $S_{\mathrm{CS}} $ defines a gauge-invariant action. Furthermore, the fact that Lorentz indices are contracted with the Levi-Civita pseudo-tensor, instead of the usual metric $g_{\mu \nu}\,$, signals that equation \eqref{eq:cs_lag} is a topological field theory, \textit{i.e.} it is a metric independent 3-form $\mathrm{A} \wedge \mathrm{d}\mathrm{A}\,$. This entails that the Chern-Simons term is invariant under coordinate transformations, and hence, the corresponding stress-energy tensor is
\begin{equation}
T_{\,\mu \nu} = - \frac{2}{\sqrt{- g}}\frac{\delta\, S_{\mathrm{CS}}}{\delta\, g^{\mu \nu}} = 0\; ,
\end{equation}
implying that the Hamiltonian associated to a Chern-Simons term vanishes identically, namely $H_{\mathrm{CS}} = 0$. A direct consequence 	is that the spectrum of such a theory is given by a number of states $\mathcal{D}$ at zero energy  which are topologically degenerate, meaning that when the system is put in a manifold $\mathcal{M}$ of genus $g$, the number of states, or degeneracy, is $\mathcal{D}=\kappa^{\,g}$. Another peculiarity is that the Euler-Lagrange equations for the Chern-Simons action \eqref{eq:cs_lag} yield $F_{\,\mu \nu} = 0 \,$, where $F_{\,\mu \nu}= \partial_{\,\mu} A_{\,\nu} - \partial_{\,\nu} A_{\,\mu}$ is the "electromagnetic" field strength tensor. Hence, solutions are trivial, meaning $A_{\mu}$ is a pure gauge or flat connection, so there are no free propagating modes for the gauge field. Notice that this is in stark contrast to the usual case of pure Maxwell's electromagnetism in which the equivalent equation is $\partial_{\,\nu} \,F^{\,\mu \nu} =0$, which has solutions in the form of plane waves. Actually, in a theory of the form $S=S_{\mathrm{Maxwell}}+S_{\mathrm{CS}}$ \cite{deser1982topomass}, photons acquire a topological mass $m_{\,\mathrm{ph}} \propto \kappa$, so a useful interpretation is that pure Chern-Simons \eqref{eq:cs_lag} is a theory of electromagnetism where "photons" become infinitely massive and cease to propagate. Thus, it is clear that with the Chern-Simons Lagrangian being first-order in derivatives, there are intriguing consequences compared to ordinary electromagnetism, which is second-order. 

\subsection{Coupling to Matter}
Let us now consider the scenario in which the Chern-Simons gauge field is coupled to a conserved current $j^{\,\mu} \equiv (\rho, \,\mathbf{j})$ representing some matter field. The total Lagrangian density will look like $\mathcal{L} = \mathcal{L}_{\mathrm{CS}} + \mathcal{L}_{\mathrm{int}}$, where $\mathcal{L}_{\mathrm{int}} = -  j^{\,\mu}A_{\mu}$. Hence, it is straight forward to compute the Euler-Lagrange equations in the usual way, 
yielding
\begin{equation}\label{eq:e_of_motion}
\frac{\kappa e^{2}}{2\pi \hbar} \, \epsilon^{\,\mu \nu \lambda} \,\partial_{\nu}\, A_{\lambda} = j^{\,\mu} \;,
\end{equation}
which is nothing but \textit{Hall's law}, where we identify the Hall conductivity as $\sigma_{H} = \kappa e^{2}/h\,$, which is quantised in units of the von-Klitzing constant. As can be seen, the equation of motion for the gauge field is non-trivial in the presence of matter. By taking the spacetime derivative $\partial_{\mu}$ on both sides we may verify that the Bianchi identity $ \epsilon^{\,\mu \nu \lambda} \partial_{\mu} F_{\nu \lambda} = 0$ is fulfilled, or equivalently, that the current is indeed conserved $\partial_{\mu} \,j^{\,\mu} = 0$. While it is clear that $A_{\mu}$ is a dynamical gauge variable, its dynamics is completely determined by the presence of a matter current $j^{\,\mu}$. We thus say that a Chern-Simons term provides a constraint telling the "electromagnetic" field to move whenever and however matter does. This is more intuitively laid out when writing equation \eqref{eq:e_of_motion} in components and by computing a simple example. Let us consider
\begin{subequations}
\begin{align}
\label{eq:subeq_1}
B = \epsilon^{\,i j} \partial_{\,i} \,A_{j} = \frac{2\pi \hbar}{\kappa e^{2}} \rho\;, \\
\label{eq:subeq_2}
\epsilon^{\,i j}  E_{j} = \epsilon^{\,i j} \big(\partial_{j}\, A_{0} - \partial_{\,t}\, A_{j}  \big) = \frac{2\pi \hbar}{\kappa e^{2}} j^{\,i}\;,
\end{align}
\end{subequations}
where we have defined the "electric" and "magnetic" fields for the gauge field $A_{\mu}$. We note that in the plane, the "magnetic" field is a pseudo-scalar, while the "electric" field is a pseudo-vector. Also, equation \eqref{eq:subeq_2} can be obtained from \eqref{eq:subeq_1} and conservation of current. Namely, by taking the time derivative of \eqref{eq:subeq_1} we obtain $ \, \partial_{\,t}\, B = 2\pi \hbar \,(\kappa e^{2})^{-1} \, \partial_{\,t} \,\rho\,$, which after substitution of the conservation law, and integration over spatial variables, yields equation \eqref{eq:subeq_2} with partial gauge-fixing $A_{0} = 0$, up to an integration constant. Thus, the relevant information is actually contained already in \eqref{eq:subeq_1}, also known as the flux-attachment condition. If we define the "magnetic" flux as $\Phi = \int d^{2}\mathbf{r} \, B\,(t, \mathbf{r})\,$, and the "charge" as $Q = \int d^{2}\mathbf{r} \, \rho \,(t, \mathbf{r})\,$, we verify that there is an explicit local equivalence $\Phi = h \,(\kappa e^{2})^{-1} Q\,$ between "charge" and  "magnetic" flux in the system. A more useful interpretation is that equation \eqref{eq:subeq_1} acts effectively both as a local constraint and as an equation of motion for the gauge field, meaning that the density $\rho$ dictates locally what is the form of the vector potential $\mathbf{A}$. A natural way to support this observation is by writing the Lagrangian density in components
\begin{equation}
\begin{split}
\mathcal{L} = A_{0}\, \bigg(\frac{\kappa e^{2}}{4\pi \hbar}B - \rho \bigg)  + \frac{\kappa e^{2}}{4\pi \hbar} \epsilon^{\,ij} A_{i}\, E_{j} - j^{\,i} A_{i}
\end{split}
\end{equation}
and noting that the component $A_{0}$ plays the role of a Lagrange multiplier enforcing the \textit{Gauss's law} \eqref{eq:subeq_1}, which we can compare with the more familiar one that would arise in Maxwell's electromagnetism, namely $\nabla \cdot \mathbf{E} = \rho\,$.
Hence, if we wished to obtain the corresponding Hamiltonian for this system, in the temporal gauge ($A_{0} = 0$), we would find
\begin{subequations}\label{eq:const_H}
\begin{align}
H = \int d^{2}\mathbf{r}\;\, \mathbf{j} \cdot \mathbf{A}\\
G \,(\mathbf{r})\ket{\mathrm{Phys}} \equiv B\,(\mathbf{r})  \ket{\mathrm{Phys}} = \frac{2\pi \hbar }{\kappa e^{2}}\, \rho\,(\mathbf{r}) \ket{\mathrm{Phys}} 
\end{align}
\end{subequations}	
which corresponds to the gauge-matter coupling in addition to a Gauss's law restricting the Hilbert space of the system to the physical states $\ket{\mathrm{Phys}}$, and such that $\,[G,H]=0\,$ at any point in space and time. Notice that the constrained Hamiltonian \eqref{eq:const_H} resembles the Hamiltonian approach to lattice gauge theories \cite{kogut1975hamiltonian, kogut1979introduction}. Thus, we see that an Abelian Chern-Simons matter theory can be thought of as a way to give restricted dynamics \footnote{Meaning that they are fully determined by the matter content of the theory.} to an otherwise background gauge field. As a last remark, it is possible to integrate out the Chern-Simons gauge field, and rewrite the Lagrangian density in terms of matter-only degrees of freedom. However, this does not come for free, since this is known to yield a Hopf term \footnote{For a Lorenz gauge $\partial_{\mu}\, A^{\mu} = 0$ this looks like $\mathcal{L_{\mathrm{Hopf}}} \propto j_{\mu} \frac{\epsilon^{\,\mu \nu \lambda} \partial_{\nu}}{\Box} j_{\lambda}\,$, see for instance \cite{zee2010quantum}. This term is also referred to in the literature as a long-range statistical interaction \cite{marino1993quantum}.}, which renders the Lagrangian nonlocal. We refer the reader to \cite{dunne1999aspects,marino1997flux} (and references therein) for further properties of Chern-Simons theory.

\section{Flux attachment}\label{sec:flux_sense}

\begin{figure}[h]%
	\centering
	\includegraphics[width=.99\linewidth]{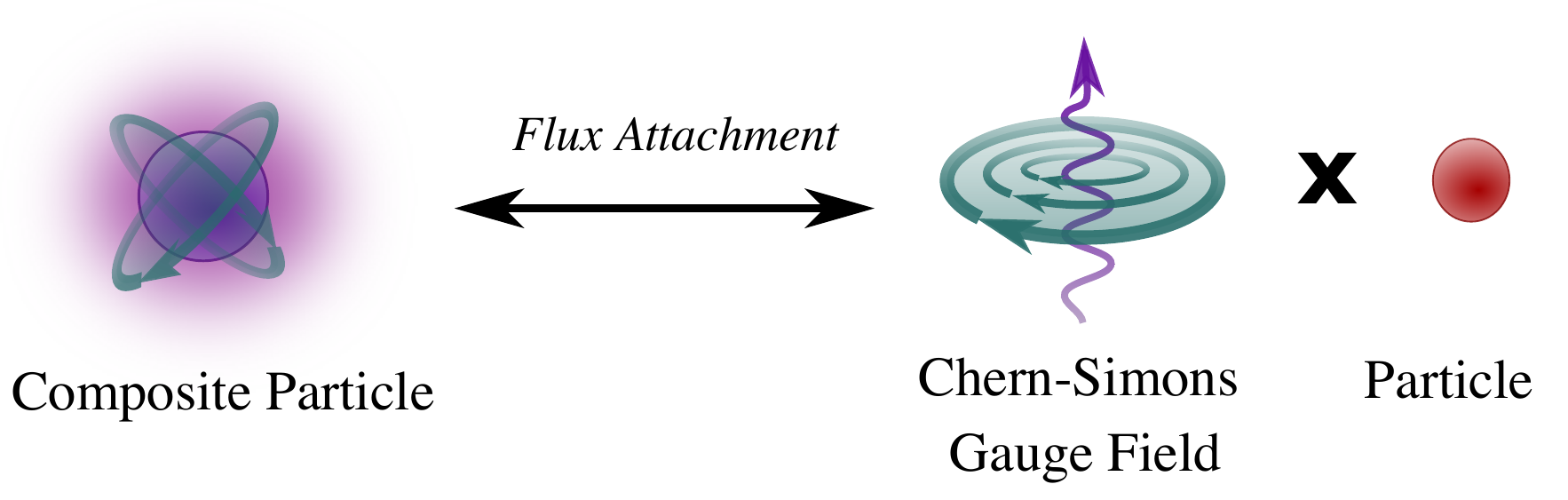}
	\caption{\textbf{Schematic of the composite particle picture.} Flux attachment is a mechanism by which charged particles capture magnetic flux quanta and become composite entities. These composites might have different properties from the bare particles, in particular they can be anyons.}%
	\label{fig:composite_part}%
\end{figure}

So far we have framed the Abelian Chern-Simons matter theory as an unusual type of gauge theory. However, an alternative, and probably more physically insightful interpretation in terms of geometric phases is possible. In 1976 A. Goldhaber \cite{goldhaber1976connection} noticed an anomalous relation of spin and statistics \footnote{See also the work of Leinaas and Myrheim \cite{leinaas1977theory}.} in charge-monopole composites. This work was revisited by Wilczek \cite{wilczek1982magnetic} and reframed as a \textit{gedankenexperiment}  in which a particle of charge $e$  in the plane orbits a solenoid (also known as a \textit{flux-tube}) placed in the transverse direction and enclosing a flux $\Phi$. In this way, a rigid bound state formed by the charged particle and the flux-tube can be seen as a single composite particle (see figure \ref{fig:composite_part}). One could then try to adiabatically transport one such charge + flux-tube composite over a closed contour $\mathcal{C}$ around a second one. The composite particle's wavefunction would then pick up an Aharonov-Bohm phase
\begin{equation}
\Psi \longrightarrow e^{\,i\, \alpha_{\mathrm{AB}}}\; \Psi = e^{\,i \frac{e}{\hbar} \Phi} \;\Psi \;.
\end{equation}
The realisation that the value of $\Phi$ defines a fractional value for the angular momentum $L_{z}$ after elimination of the gauge potential via a singular gauge transformation, led Wilczek to define the notion of an \textit{anyon} \cite{wilczek1982quantum,arovas1985statistical} as a particle-flux composite. This means that, upon exchanging two composites, the total wavefunction can acquire a general phase shift. This is easily illustrated by taking the flux attachment relation \eqref{eq:subeq_1}, and realising that for a point particle $\rho =e \,\delta^{\,(2)}(\mathbf{r})\,$ the Aharonov-Bohm phase for a full winding is $2\pi \kappa^{-1}$, so that for an exchange the phase factor is $e^{\,\pm i \,\pi \kappa^{-1}}$. The $+$ ($-$) sign denotes anticlockwise (clockwise) exchange, and the Chern-Simons level $\kappa$ can take arbitrary values.

This idea was then linked to the nonlinear $\sigma$-model \cite{wilczek1983linking}, used in the context of resonance-valence-bond states \cite{kalmeyer1987equivalence}, and finally reintroduced by Jain \cite{jain1989composite} in the context of the fractional QHE understood as an integer QHE of composite particles, defined as electrons "dressed" with flux-tubes. This "dressing" is what we mean by flux attachment in this context \footnote{This is also discussed by Polyakov \cite{polyakov1988fermi} as a boson-fermion transmutation on the plane.}. More formally, it means performing a singular gauge transformation to the wavefunction of the system. The immediate effect of such a transformation is the introduction (or removal) of a minimally coupled singular vector potential, often referred to as the \textit{statistical} gauge field (see Appendix \ref{sec:bosonization} for further details). In a nutshell, by attaching flux-tubes to particles one can transform a strongly-correlated problem of electrons into a weakly-correlated problem of composite particles. Macroscopic descriptions \cite{girvin1987off,zhang1989effective,read1989order,lopez1991fractional} of the fractional QHE rely on the appearance of such singular potential as a Chern-Simons gauge field. The explicit form of which can be derived from solving the flux attachment condition \eqref{eq:subeq_1}, giving rise to
\begin{equation}\label{eq:gauge_arbitrary}
A^{\,i}(t,\mathbf{r}) = \partial^{\,i} \Lambda\,(t,\mathbf{r}) + \frac{\hbar}{\kappa e^{2}} \,\epsilon^{\,ij}\int d^{2}\mathbf{r'}\; G_{\,j\,}(\mathbf{r} - \mathbf{r'}) \, \rho\,(t,\mathbf{r'}) \;,
\end{equation}
where $\Lambda$ is an arbitrary gauge. Note that the Green's function renders $\mathbf{A}$ a \textit{singular} pure gauge such that $\mathbf{G}\,(\mathbf{r}) = \nabla \varphi \,(\mathbf{r})\,$, with $\varphi$ being the polar angle. What we mean by this is that $\mathbf{A}$ is a local, although not global, pure gauge, provided the function $\varphi = \tan^{-1}(y/x)$ is multivalued. This implies that $\epsilon^{\,ij}\partial_{i} \partial_{j} \varphi = 2\pi \delta (\mathbf{r})$ (see Appendix \ref{sec:multi}), and hence, the Green's function is a vortex. In the Coulomb gauge $\nabla\cdot \mathbf{A} = 0$, and equation \eqref{eq:gauge_arbitrary} simplifies to
\begin{equation}\label{eq:vect_pot_form}
\mathbf{A}\,(t,\mathbf{r}) = \frac{\hbar}{\kappa e^{2}} \,\bigg[\,\hat{z} \times \int d^{2}\mathbf{r'}\; \frac{\mathbf{r}-\mathbf{r'}}{|\mathbf{r}-\mathbf{r'}|^{2}} \, \rho\,(t,\mathbf{r'}) \,\bigg]\;.
\end{equation}
This allows for a powerful reinterpretation of the fractional QHE as an emergent Chern-Simons matter theory at low energies, where the singular gauge potential appears in a collective rearrangement of the planar electron gas under the influence of a strong transverse external magnetic field. We thus see the manifestation of two sides of the same coin, we can describe the system as a gas of particle-flux-tube composites or as a problem of physical particles subject to a Chern-Simons gauge field. This unified view was crucial for the modern interpretation of the fractional QHE in the half-filled Landau level \cite{halperin1993theory} and the recent discovery of Dirac composite fermions \cite{son2015composite,son2018dirac}.

\section{From Chern-Simons to Ultracold Gases}\label{sec:to_qsim}

As we have highlighted already, the Chern-Simons term is a topological field theory, with a vanishing Hamiltonian, that can be thought of as a local constraint fixing the form of the gauge field. We call this constraint flux attachment. We further notice that the vector potential \eqref{eq:vect_pot_form} depends on density $\rho$ and has a vortex kernel, which for a point particle is nothing but that of the usual Aharonov-Bohm effect.

 Within this framework, an obvious question to ask is how does such a Chern-Simons term appear in the first place? As a matter of fact, such a contribution can be radiatively induced in Quantum Electrodynamics (QED) \cite{dunne1999aspects} or understood from \textit{pseudo-QED} descriptions \cite{marino1993quantum}. Yet, the above mechanisms correspond to the effective macroscopic picture of a condensed matter system, and they do not offer a physical explanation of the microscopic origin of the Chern-Simons gauge field or topological order. In fact, the Chern-Simons term is often added "by hand" and regarded as \textit{emergent} in the low energy effective theory. That is, it appears phenomenologically as a collective rearrangement of the many-body system. 

In the following we approach the question in reverse and ask, how can we engineer a Chern-Simons term starting from a microscopic system? The key point that allows us to do so is precisely realising that the information about Chern-Simons is already contained in equation \eqref{eq:vect_pot_form}, which is ensured by flux attachment. The main challenge is then how to induce such a pinning of flux. We argue that this can be achieved for a charge-neutral system by making use of artificial gauge fields in which a carefully designed Berry connection term plays the role of the effective Chern-Simons gauge field. Hence, starting from a microscopic many-body Hamiltonian, we aim to recover an Abelian Chern-Simons + matter theory in an effective macroscopic description of such system. Notice that so far we have not specified the form of the matter component so far, so the discussion above remains completely general regardless of the system or platform. 

In view of a potential realisation with the current state-of-the-art experimental techniques in atomic physics, we focus on the case of a Bose-Einstein condensate. While our protocol is inevitably idealised and approximate, it provides a proof-of-concept scheme. In Bose-Einstein condensates of dilute atomic gases, the dominant interaction is typically that of molecular potentials, namely hard-core repulsion as $r \rightarrow 0$ with an attractive van der Waals tail $\sim 1/r^{6}$. Normally, this can be described by a $\delta$-function type interaction for s-wave scattering, where the average interparticle distance is $0.1 \mathrm{\mu m}$ for a condensate of $n=10^{14}$ $\mathrm{atoms/cm^{3}}$, see \textit{e.g.} \cite{inguscio2013atomic}. Furthermore, the scattering length $a$, characterising the strength of the interaction, can in general be tuned as a function of an external field near Feshbach resonances.

\section{The non-linear gauge potential} \label{sec:dens_dep_bec}
 Let us consider a system consisting of $N$ two-level bosonic atoms where the internal states $\ket{1}$ and $\ket{2}$ are coherently coupled by a laser beam, and the atoms interact pairwise. The many-body Hamiltonian describing this system \cite{dalibard2011colloquium,dd2015introduction} in the rotating-wave approximation is 
\begin{equation}\label{eq:two_level_ham}
H = \sum_{n=1}^{N} \bigg(\frac{\mathbf{p}_{n}^{2}}{2M} + V_{n}+ U_{n} \bigg) \otimes \mathbb{I}_{\mathcal{H} \backslash n} + \sum_{n<m}^{N} \mathcal{V}_{n m} \otimes \mathbb{I}_{\mathcal{H} \backslash \{n,m\}} \;, 
\end{equation}
where $\mathbf{p} \equiv - i \hbar \nabla$ is the momentum operator, $V_{n}$ is an external (\textit{e.g.} confining) potential, and the identity matrices simply provide the correct dimensionality for the Hamiltonian. The light-matter coupling matrix is
\begin{equation}\label{eq:coupling_matrix}
U_{n} = \frac{\hbar }{2} \begin{pmatrix}
\Delta & \Gamma^{\,*} \\
\Gamma & - \Delta
\end{pmatrix} \;,
\end{equation}
where $\Delta = \omega_{L} -\omega_{A}$ is the detuning between the laser and atomic transition frequencies which can be a function of the atomic centre-of-mass position. The Rabi frequency $\Gamma = |\Gamma|\,e^{i\phi} = (\mathbf{d}_{12}\cdot \mathbf{E}_{0})\,\hbar^{-1}$ characterises the strength of the light-matter interaction. 
Introducing the notation
\begin{equation}
\Omega = \sqrt{\Delta^{2} + |\Gamma|^{2}} \; , \;\;\;\;\;\;  \cos \theta =\frac{\Delta}{\Omega}    \; ,\;\; \;\;\;\; \sin \theta = \frac{|\Gamma|}{\Omega} \;  , 
\end{equation}
allows us to redefine variables in terms of the mixing angle $\theta \equiv \tan^{-1} (|\Gamma|/\Delta)$, the generalised Rabi frequency $\Omega$, and the laser phase $\phi$. The light-matter coupling matrix in equation \eqref{eq:coupling_matrix} then becomes
\begin{equation}
U_{n} =  \frac{\hbar \Omega}{2} \begin{pmatrix}
\cos \theta & e^{-i \phi } \sin \theta \\
e^{i \phi} \sin \theta & - \cos\theta
\end{pmatrix} = \frac{\hbar \Omega}{2} \, \mathbf{n}\cdot \bm{\sigma}\; ,
\end{equation}
where we re-expressed $U_{n}$ as the product of a unit vector $\mathbf{n}$ characterized by angles $\theta$ and $\phi$, and a vector of Pauli matrices $\bm{\sigma}$. In fact, this is just the spin-$1/2$ Berry phase problem. The eigenstates are given by
\begin{equation}\label{eq:dressed_states}
\ket{\chi ^{(0)}_{+}} = \begin{pmatrix}
\cos\, (\theta/2) \\
e^{i \phi} \sin\, (\theta/2)
\end{pmatrix} , \;\; \ket{\chi ^{(0)}_{-}} = \begin{pmatrix}
\sin \,(\theta/2) \\
- e^{i \phi}\cos \,(\theta / 2)
\end{pmatrix} 
\end{equation}
with corresponding eigenvalues $\varepsilon^{(0)}_{\pm} = \pm \frac{\hbar \Omega}{2}$. One can show that \eqref{eq:dressed_states} forms an orthonormal set of vectors $\{\ket{\chi^{(0)}_{j}}\}$ \footnote{We introduce the superscript $(...)^{\,(0)}$ in order to avoid confusion at later stages, when we add a perturbation to $U$.} with $j=\{+,-\}$, which will be used as the basis for the \textit{internal} Hilbert space. In the context of quantum optics these states are commonly known as \textit{dressed states}. 

\begin{figure}[h]%
	\centering
	\includegraphics[width=.99\linewidth]{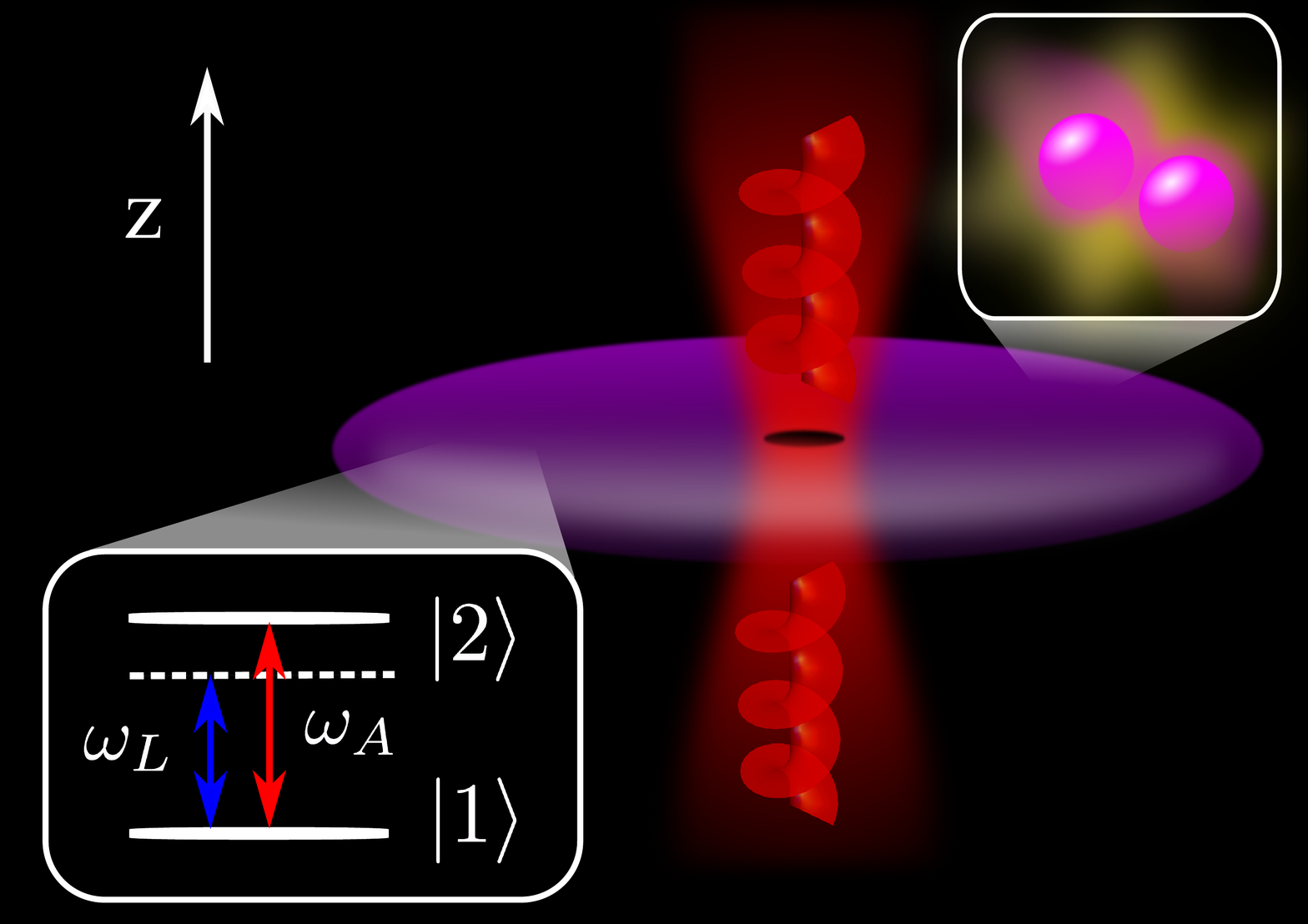}
	\caption{\textbf{Coherently coupled Bose-Einstein condensate in 2+1 dimensions.} Atoms have a two-level internal structure (lower inset) and interact pairwise (upper inset). A laser beam with orbital angular momentum imprints a localised density profile and effectively generates flux attachment.}%
	\label{fig:condensat_fig}%
\end{figure}

The interaction term in equation \eqref{eq:two_level_ham} has the form
\begin{equation}
\mathcal{V}_{n m} = \mathrm{diag}\,\big[\,g_{11},\,g_{12},\,g_{12},\,g_{22}\,\big] \, \mathcal{K}\,(\mathbf{r}_{n}-\mathbf{r}_{m}) \;,
\end{equation}
where $\mathcal{K}\,$ is an arbitrary two-body interaction, which in the limit of zero-range interactions is $\mathcal{K}\,(\mathbf{r}_{n}-\mathbf{r}_{m}) \rightarrow \delta\,(\mathbf{r}_{n}-\mathbf{r}_{m})$ with coupling constants $g_{ij} = 4\pi \hbar^{2} a_{ij} M^{-1}$ \footnote{Note that in 2+1 dimensions, the coupling constant is slightly modified as a result of confinement of the system in the $z$ axis, and hence $g_{2+1} = g_{3+1} \,\big(a_{z} \sqrt{2 \pi}\big)^{-1}$, see \textit{e.g.} \cite{salasnich2002effective}.} characterising the strength of the interactions in terms of the scattering lengths $a_{ij}$ for three different channels. The indices $i,j=1,2$ label the two internal states of the atom, see figure \ref{fig:condensat_fig}. In the following, we extend the treatment of Refs. \cite{edmonds2015elementary, butera2016vortex, butera2018curved, goldman2014light} to include long-range interactions and non-zero detuning. 

\subsection{Mean-field Approximation and Expansion}

Typical number densities in Bose-Einstein condensates are $\rho \sim 10^{13} -10^{15}\,\mathrm{cm}^{-3}$. These dilute conditions correspond to a weakly interacting regime. Thus, it is sensible to consider a mean-field (MF) variational ansatz for the many-body wavefunction as the symmetrised product of single-particle wavefunctions $\Psi \,(\mathbf{r}_{1},\mathbf{r}_{2},\dots,\mathbf{r}_{N}) = \prod_{l=1}^{N} \tilde{\psi} \,(\mathbf{r}_{i})$, satisfying the normalisation $\int d^{2}\mathbf{r} \, | \tilde{\psi}\, (\mathbf{r})|^{2}= 1$. We can define an order parameter acting as a condensate wavefunction $\psi \,(\mathbf{r}) = \sqrt{N}\, \tilde{\psi}\,(\mathbf{r})$. In this limit, the energy-scales corresponding to mean-field interparticle interactions are much smaller than those of the light-matter coupling, meaning $g_{ij} \rho_{j} \ll \hbar \Omega$ with $\rho_{j} = |\psi_{j}|^{2}$. Then, the interparticle interaction term reads
\begin{equation} \label{eq:mf_interaction}
\mathcal{V}_{\,\mathrm{MF}} = \frac{1}{2} \begin{pmatrix}
\nu_{1} & 0 \\
0 & \nu_{2}
\end{pmatrix} \, ,
\end{equation}
where $\nu_{i} = \sum_{j=1}^{2} g_{ij} \int d^{2}\mathbf{r} \;\mathcal{K}\,(\mathbf{r} - \mathbf{r'})\,\rho_{j}\,(\mathbf{r'})$ acts as an effective mean-field interaction-induced detuning between atomic levels. This enables us to treat $\mathcal{V}_{\mathrm{\,MF}}\,$ as a small perturbation of the laser-atom coupling. We thus write the first-order perturbed dressed states and energies as
\begin{equation}
	\ket{\chi_{\pm}(\mathbf{r})} \approx \ket{\chi^{(0)}_{\pm}(\mathbf{r})} + \ket{\chi^{(1)}_{\pm}(\mathbf{r})} , \;\;\;\; \varepsilon_{\pm} \approx \varepsilon^{(0)}_{\pm} + \varepsilon^{(1)}_{\pm} .
\end{equation} 
The unperturbed states are given by equation \eqref{eq:dressed_states}, while corrections to the eigenstates are
\begin{equation}
\ket{\chi^{(1)}_{\pm}(\mathbf{r})} = \pm \frac{\sin\, (\theta)}{4\hbar \Omega} (\nu_{1} - \nu_{2}) \ket{\chi^{(0)}_{\mp}(\mathbf{r})} \, ,
\end{equation}
with eigenvalues
\begin{subequations}
\begin{align}
\varepsilon^{(1)}_{+} &= \frac{1}{2} \Big( \nu_{1} \cos^{2}(\theta/2) + \nu_{2} \sin^{2}(\theta/2)  \Big)\;, \\
\varepsilon^{(1)}_{-} &= \frac{1}{2} \Big( \nu_{1} \sin^{2}(\theta/2) + \nu_{2} \cos^{2}(\theta/2)  \Big) \;.
\end{align}
\end{subequations}

We now write the full state vector for the two-level condensate as a linear combination of the perturbed dressed states
\begin{equation}\label{eq:eigen_decomposition}
\ket{\Psi \,(t,\mathbf{r})} = \sum_{j=\,(+,-)} \psi_{j}\,(t,\mathbf{r}) \ket{\chi_{j}(\mathbf{r})} \; ,
\end{equation}
so that the dressed states are steady-state solutions and coefficients $\psi_{j}$ contain the temporal dependence.

\subsection{Adiabatic Approximation and Effective Model}

We would like to compute the effective action for the condensate. To proceed, we will rely on the adiabatic approximation, meaning that when the system is prepared in a given eigenstate $\ket{\chi_{\pm}(\mathbf{r})}$, it will remain in this state at any given time. In view of the above, we can project the problem on the subregion of its Hilbert space in which the system is initially prepared. This implies that if the system is prepared in the $\ket{\chi_{\pm}(\mathbf{r})}$ dressed state, the coefficient $\psi_{\mp}\,(t,\mathbf{r}) \approx 0$ for any $t$. Thus, we obtain the mean-field 
Hamiltonian
\begin{equation}
H_{_{\mathrm{MF}}} = \frac{\mathbf{p}^{2}}{2M} \otimes \mathbb{I}_{2} + V \,(\mathbf{r})+ U \, (\mathbf{r})+ \mathcal{V}_{\mathrm{\,MF}} \;.
\end{equation}
After projection of the system onto one of its ($\pm$) dressed states, the effective model becomes $i\hbar \,\partial_{t} \,\psi_{\pm}= H^{\,\mathrm{eff}} _{_{\mathrm{\pm}}} \,\psi_{\pm}$,
where
\begin{equation} \label{eq:projected_ham}
H^{\,\mathrm{eff}} _{_{\mathrm{\pm}}} \approx \frac{\big(\mathbf{p}-\bm{\mathcal{A}}_{\,\pm} \big)^{2}}{2M} + V(\mathbf{r}) + \mathcal{W}_{\mp \pm} + \varepsilon^{(0)}_{\pm} + \varepsilon^{(1)}_{\pm} \; ,
\end{equation} 
where $\bm{\mathcal{A}}_{\,\pm} = i\hbar \braket{\chi_{\pm} | \nabla \chi_{\pm}} \approx \mathbf{A}_{\pm} + \bm{a}_{\pm}$ has the form of a Berry-connection term, which plays the role of a minimally-coupled \textit{synthetic} vector potential. More explicitly, 
\begin{equation}
\mathbf{A}_{\pm} = i\hbar \braket{\chi^{(0)}_{\pm} | \nabla \chi^{(0)}_{\pm}} = \pm \frac{\hbar}{2} \big[ \cos\,(\theta) -1 \big] \nabla \phi
\end{equation}
corresponds to the single-particle contribution, while
\begin{flalign}
\bm{a}_{\pm} &= i\hbar \, \Big[\braket{\chi^{(0)}_{\pm} | \nabla \chi^{(1)}_{\pm}} + \braket{\chi^{(1)}_{\pm} | \nabla \chi^{(0)}_{\pm}} \Big] \\
 &= \pm \frac{\sin^{2}(\theta)}{4\Omega} \big(\nu_{1} - \nu_{2} \big) \nabla \phi
\end{flalign}
is the first-order correction induced by interactions.
Similarly, $\mathcal{W}_{\mp \pm} = \frac{\hbar^{2}}{2M}| \braket{\chi^{(0)}_{\mp} | \nabla \chi^{(0)}_{\pm}}|^{2}$ is a synthetic geometric scalar potential. It is worth noting that the many-body information in the projected mean-field Hamiltonian \eqref{eq:projected_ham} is contained in the effective interaction-induced detunings $\nu_{i}$, which in turn are functions of the density $\rho_{i}$. Introducing the matter density in the dressed state basis $\rho_{\pm} = |\psi_{\pm}|^{2}$ \footnote{Notice the difference between $\rho_{i}$, expressed in the original basis $\{\ket{1}, \ket{2}\}$, and $\rho_{\pm}$, defined in the dressed state basis $\{\ket{-}, \ket{+}\}$.}, we can explicitly see this dependence on interactions in
\begin{equation}\label{eq:dens_dep_gauge}
\bm{a}_{\pm} = \pm \frac{f_{\pm}(\theta)}{8 \Omega} \Big[ \int d^{2}\mathbf{r'} \,\mathcal{K}(\mathbf{r} - \mathbf{r'})\rho_{\pm}(\mathbf{r'}) \Big]\,\nabla\phi \equiv F\,(\mathbf{r})\, \nabla\phi \,,
\end{equation}
which is an interaction-dependent synthetic gauge potential. The explicit form of $f$ is
\begin{equation}
\begin{split}
 f_{\pm} (\theta) &=  \pm \sin^{2}(\theta) \, \Big[4 g \cos\,(\theta) \pm (g_{11} - g_{22})\Big] \;,
\end{split}
\end{equation}

 where we have defined $g \equiv (g_{11}+g_{22}-2 g_{12})/4$. Notice that at zero detuning and for contact interactions, we recover the results from \cite{edmonds2013simulating}. 
\subsection{Finding Synthetic Flux Attachment}
Defining the total magnetic field as $\bm{\mathcal{B}}_{\pm} = \mathbf{B}_{\pm} + \mathbf{b}_{\pm}$, what remains now is showing that the magnetic field associated with equation \eqref{eq:dens_dep_gauge}, namely
\begin{equation}\label{eq:magn_field}
	\mathbf{b}_{\pm} = \nabla F(\mathbf{r}) \times \nabla\phi + F(\mathbf{r})  \, \nabla \times \nabla \phi \; ,
\end{equation}
can represent flux attachment. It is tempting to try to find an interaction kernel for which a simple choice of the laser phase would yield equation \eqref{eq:subeq_1}. The kernel needed would require long-range interactions $\sim 1/r$ in addition to a vortex-like structure (see discussion in Appendix \ref{sec:kernel}). However, from an implementation point of view, it would be desirable that interactions remain short-ranged meaning that the interaction kernel becomes a delta function, \textit{i.e.} $\,\mathcal{K}\,(\mathbf{r}) \cong \delta \,(\mathbf{r})\,$. The latter implies constraining light-matter coupling parameters $\theta$, $\phi$ and $\Omega$. 

\begin{figure}[h]%
	\centering
	\includegraphics[width=.99\linewidth]{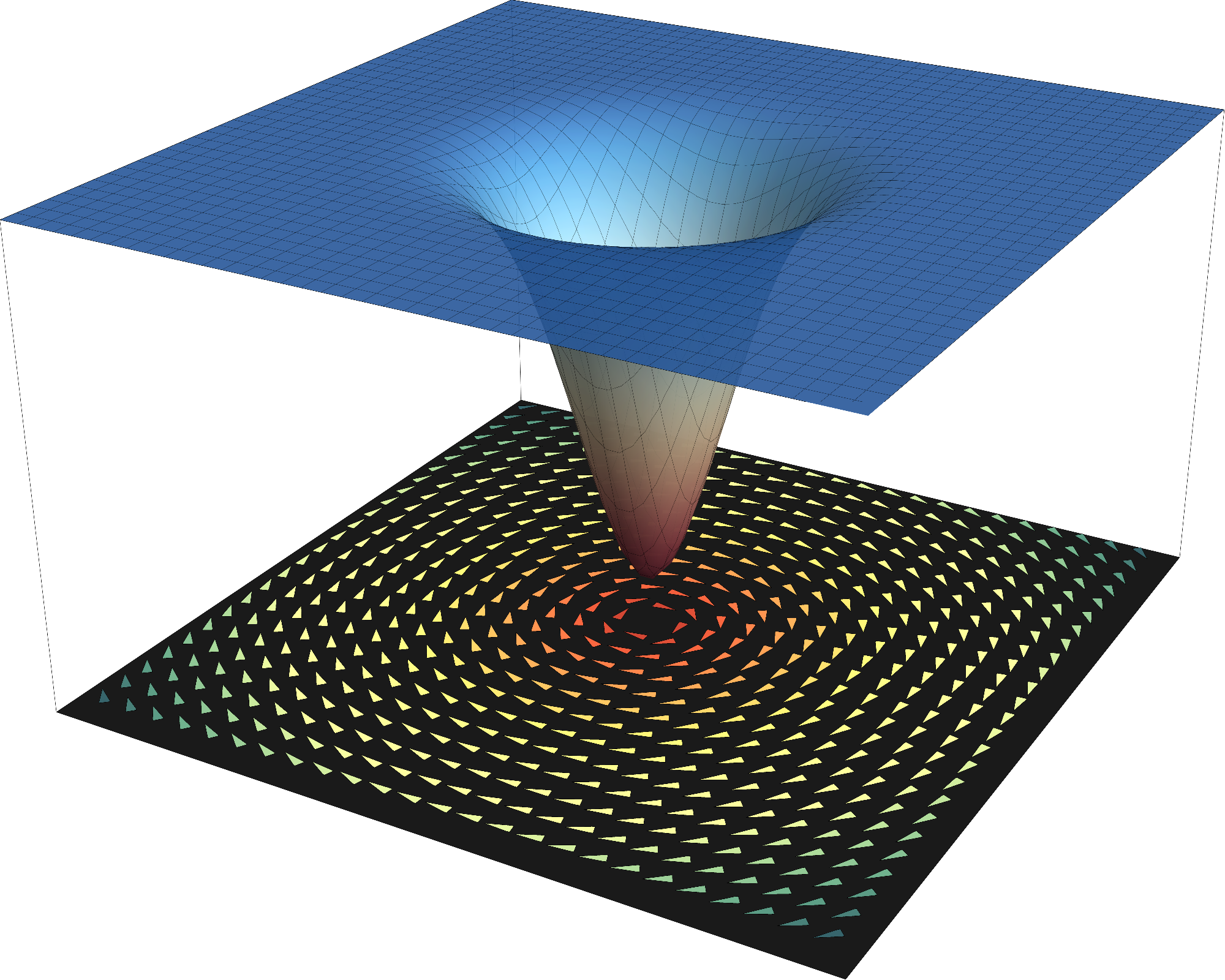}
	\caption{\textbf{Depiction of synthetic flux attachment.} A localised density profile of the condensate on the $x-y$ plane (upper), and the corresponding vortex shape of the vector potential $\bm{a}_{\pm}$ (lower). Colour coding on the contour plot depicts a radial decay as $\sim 1/r\,$ for the vector potential on top of the modulation by matter density. Flux attachment ensures that the density profile is proportional to the synthetic magnetic field.}%
	\label{fig:flux_fig}%
\end{figure}

 Let us choose a laser beam with orbital angular momentum (\textit{e.g.} Laguerre-Gaussian mode) so that $\phi = l \varphi$, where $l$ is the winding number and $\varphi$ is the polar coordinate in the plane. Assuming now a rotationally symmetric density profile $\rho$, mixing angle $\theta$, and generalised Rabi frequency $\Omega$, we are left with
\begin{equation}\label{eq:magn_field_exp}
\mathbf{b}_{\pm} = \pm\frac{l}{r} \Bigg[ \rho_{\pm}\,\bigg( 2 \pi r  \delta^{ \,(2)}( \mathbf{r}) \frac{ f_{\pm}(\theta) }{8 \Omega} + \partial_{r} \frac{ f_{\pm}(\theta) }{8 \Omega}\bigg) + \frac{ f_{\pm}(\theta) }{8 \Omega} \partial_{r} \rho_{\pm}  \Bigg] \,\hat{z}\;.
\end{equation}
From equation \eqref{eq:magn_field_exp} we see that two constraints can be identified when comparing with \eqref{eq:subeq_1} at $r \neq 0\,$. The first one is
\begin{equation}\label{eq:const1}
\frac{l}{r} \,\bigg(\partial_{r} \frac{ f_{\pm}(\theta) }{8 \Omega}\bigg) = \frac{2\pi \hbar}{\kappa}\;,
\end{equation}
which fixes the form of $ f_{\pm}(\theta)\,(8 \Omega)^{-1}$. In addition to equation \eqref{eq:const1}, we also require
\begin{equation}\label{eq:const2}
\rho_{\pm}\,\bigg(\partial_{r} \frac{ f_{\pm}(\theta) }{8 \Omega}\bigg) \gg \frac{ f_{\pm}(\theta) }{8 \Omega} \big(\partial_{r} \rho_{\pm} \big) \;.
\end{equation}
In particular, we can consider $\rho_{\pm}$  to be sufficiently slowly varying so that its derivatives are small. This is valid for certain localised density profiles (\textit{e.g.} a Gaussian dip or a vortex, see figure \ref{fig:flux_fig}). Alternatively, this second constraint can also be seen as an "effective range" of flux attachment. Provided conditions \eqref{eq:const1} and \eqref{eq:const2} are satisfied, our system is effectively described by the Hamiltonian \eqref{eq:projected_ham} constrained by both current conservation $\partial_{\mu}\, j^{\,\mu} = 0$ and flux attachment 
\begin{equation} \label{eq:magn1}
\mathbf{b}_{\pm} \approx \pm\, \Big[\frac{2\pi \hbar}{\kappa} \rho_{\pm} +\delta^{ \,(2)}( \mathbf{r})\, \frac{2\pi l f_{\pm}(\theta)}{8 \kappa \Omega} \rho_{\pm} + \mathcal{O} \,(\partial_{r} \rho_{\pm})\Big] \,\hat{z}\, ,
\end{equation}
where the last term indicates corrections depending on the density profile. In the same vein, the single-particle magnetic field will be 
\begin{equation}\label{eq:magn2}
\mathbf{B}_{\pm} = \pm \frac{\hbar}{2} \, \Big[ - \frac{l}{r} \sin\,(\theta) \, \partial_{r} \theta + \Big(\cos\,(\theta) - 1 \Big) \, \delta^{ \,(2)}( \mathbf{r})	\Big]  \,\hat{z} \, .
\end{equation}

The Aharonov-Bohm contribution to the magnetic fields yields a non-zero magnetic field at $r = 0\,$, \textit{i.e.} that of an infinitely thin solenoid. Provided that magnetic fields in equations \eqref{eq:magn1} and \eqref{eq:magn2} have a single component, they are effectively pseudo-scalar fields, as expected.

\subsection{Recovering the Chern-Simons Term}	
We can incorporate the interacting contribution of the synthetic magnetic field through a Lagrange multiplier, and compute the effective Lagrangian density for the condensate. Considering $\mathbf{b}_{\pm} = b_{\pm}\,\hat{z}\,$, the effective description is given by
\begin{multline}\label{eq:naive_eff_L}
\mathcal{L}^{\mathrm{\, eff}}_{\pm} \approx -\frac{\kappa}{2\pi \hbar} \,a_{0}\,b_{\pm} + i\hbar \, \psi^{*}_{\pm} D_{t}\, \psi_{\pm} - \frac{\hbar^{2}}{2M} \big|\mathbf{D}\, \psi_{\pm} \big|^{2} \\
- \frac{g}{2} \big( \psi^{*}_{\pm} \psi_{\pm} \big)^{2} - \Big(V \pm \frac{\hbar \Omega}{2} +\mathcal{W}_{\mp \pm} \Big)\, \psi^{*}_{\pm} \psi_{\pm}  \;,
\end{multline}
where the field $a_{0}$ is added as the Lagrange multiplier field that introduces the constraint. Here, the condensate minimally couples to gauge fields through the gauge covariant derivative $D_{\mu} \equiv \partial_{\mu} - i \hbar^{-1}\mathcal{A}_{\mu}\,$. We have already seen that the preservation over time of the flux-attachment condition has a counterpart in terms of  an "electric" field and a current. This condition can also be incorporated into the Lagrangian using the conservation of the latter. The first term becomes nothing but the Chern-Simons term. Let us drop the dressed state subindex $\pm$ and take the time derivative of the flux attachment condition \eqref{eq:subeq_1}, giving rise to
\begin{equation}
	\partial_{\,t}\,b = \frac{2\pi \hbar}{\kappa} \partial_{\,t}\,\rho = - \frac{2\pi \hbar}{\kappa} \partial_{\,i}\,j^{\,i} \, ,
\end{equation}
where in the last step we have used the continuity equation $\partial_{\mu} \,j^{\,\mu} = 0$. After reordering and expressing the magnetic field in terms of the vector potential, we realise that
\begin{equation}\label{eq:deriv_current}
- \partial_{\,i}\,  \epsilon^{\,ij} \partial_{\,t}\,a_{j}  = \frac{2\pi \hbar}{\kappa}\,\partial_{\,i}\, j^{\,i}_{\parallel} \;,
\end{equation}
where we have used the Helmholtz decomposition of the current in parallel $\parallel$ and transverse $\perp$ components, meaning that $j^{\,i} = j^{\,i}_{\parallel} + j^{\,i}_{\perp}$. Since $j^{\,i}_{\perp} = - \epsilon^{\,ij}\partial_{\,j} \,\chi_{\perp}$, where $\chi_{\perp}$ is an unspecified function, we trivially observe that $\partial_{\,i}\, \epsilon^{\,ij}\partial_{\,j} \, \chi_{\perp} = 0\,$, and thus, $\partial_{\,i}\,j^{\,i} = \partial_{\,i}\,j^{\,i}_{\parallel}\,$. Integration of equation \eqref{eq:deriv_current} yields
\begin{equation}
\epsilon^{\,ij} \,\Bigg(\frac{2\pi \hbar}{\kappa}\partial_{j}\, \chi_{\perp} - \partial_{\,t}\,a_{j} \Bigg) = \frac{2\pi \hbar}{\kappa} j^{\,i}_{\parallel}\,.
\end{equation}
Upon identification of $a_{0} = 2\pi \hbar \kappa^{-1}\chi_{\perp}$, we conclude that
\begin{equation}
\epsilon^{\,ij} \,\mathcal{F}_{j\,0}= \epsilon^{\,ij} \,\mathcal{E}_{j} = \frac{2\pi \hbar}{\kappa} j^{\,i}_{\parallel}\,,
\end{equation}
where $\mathcal{F}_{\mu \nu}$ is the synthetic electromagnetic field strength tensor, and $\mathcal{E}$ is the synthetic electric field. Including this constraint in the Lagrangian formalism yields the Chern-Simons term, so that the effective action is
\begin{multline}\label{eq:final_eff_S}
S^{\mathrm{\, eff}}_{\pm} = \int dt\,d^{2}\mathbf{r}\; \bigg[-\frac{\kappa}{4\pi \hbar} \epsilon^{\,\mu \nu \lambda} \,a_{\mu}\,\partial_{\nu}\, a_{\lambda} + i\hbar \, \psi^{*}_{\pm} D_{t}\, \psi_{\pm} \\
- \frac{\hbar^{2}}{2M} \big|\mathbf{D}\, \psi_{\pm} \big|^{2} - \frac{g}{2} \big( \psi^{*}_{\pm} \psi_{\pm} \big)^{2} - \Big(V \pm \frac{\hbar \Omega}{2} +\mathcal{W}_{\mp \pm} \Big)\, \psi^{*}_{\pm} \psi_{\pm} \bigg] \;.
\end{multline}
Alternatively, we can argue that the flux attachment condition by itself yields the Chern-Simons term evaluated in the Coulomb gauge \cite{zhang1989effective, altland2010condensed}, meaning that the vector potential has only a transverse $\perp$ component. However, the usual covariant form of the Chern-Simons term incorporates also its parallel component. Reversing the usual Faddeev-Popov gauge fixing procedure \cite{faddeev2016feynman} reintroduces the full gauge phase space. An additional remark is that the procedure described in this section is similar to that found in the Schwinger model when eliminating the gauge field using the corresponding Gauss's law, which yields an integration constant that is used to define the so-called $\theta$ angle \cite{coleman1976more}. 

\section{Consequences of flux attachment} \label{sec:consequences}
Equation \eqref{eq:final_eff_S} provides a mean-field description of the laser-coupled Bose-Einstein condensate. More generally, this emergent effective description is that of an interacting charged superfluid minimally coupled to an internal (dynamical) Chern-Simons gauge field $a^{\,\mu}$ and an external (background) gauge field $A^{\,\mu}$. This is known as the Zhang-Hansson-Kivelson (ZHK) model \cite{zhang1989effective, read1989order}, and provides a bosonic macroscopic description of the fractional QHE in the spirit of a Ginzburg-Landau theory. In the absence of the external field $\mathbf{A}$ the system reduces to the so-called Jackiw-Pi model \cite{jackiw1990classical, jackiw1990soliton, jackiw1992self}, which can be analytically solved in the self-dual static limit, yielding multi-vortex solutions. Taubes' theorem \cite{jaffe1980vortices} guarantees that vortex solutions also exist for the ZHK model giving rise to the Chern-Simons (flux-attached) vortices \footnote{We note here that, while non-relativistic Jackiw-Pi Chern-Simons vortices are non-topological, ZHK are. Furthermore, relativistic, non-Abelian versions (and deformations) of such models are also known for yielding different families of topological and non-topological vortex solutions.} whose explicit form can also be computed, where these are akin to the well-known Abrikosov-Nielsen-Olesen vortices in type II superconductors if the dynamical gauge field were Maxwell-like. A key feature of these Chern-Simons vortices is their composite nature, \textit{i.e.} they are dyonic objects that play the role of Laughlin's (anyonic) quasiparticles carrying both electric charge and magnetic flux. These and other features follow from the effective model \eqref{eq:final_eff_S} and they are discussed in detail in the seminal work of Zhang \cite{zhang1995chern}. We highlight some of them in what follows, where it is worth identifying electric current in the charged superfluid as matter flow in our system, and charge density corresponding to matter number density.\\

\paragraph{Quantisation of the transverse flow.} 
The immediate consequence of flux attachment is that the "atomic Hall conductivity" $\sigma_{H}$ must be quantised because of topological arguments, \textit{i.e.} index theorems. This would appear in the form of clear plateaus in the Hall response, so a transport measurement is typically needed as a probe. We can imagine the creation of a tilt in the condensate in such a way that a matter current is generated \cite{leblanc2012observation}. Then, the atomic transverse response is parametrised by $\mathbf{j} = \sigma_{H} \, \nabla_{\perp}V$, where $V$ is an external (\textit{i.e.} tilting) potential. The Chern-Simons level $\kappa$ plays the role of the Landau level filling fraction $\nu \equiv \sigma_{H} / \sigma_{0} \,$, where $\sigma_{0} = (2\pi \hbar)^{-1}$. For Laughlin-like fractions, one expects $\kappa \equiv\nu = 1/m$ for $m \in \mathbb{Z}\,$.\\

\paragraph{Vortex exchange and statistics.}
As we highlighted in section \ref{sec:flux_sense}, assuming the density profiles correspond to Chern-Simons vortices, the Aharonov-Bohm phase associated to interchanging two such composites can alter the statistics of the object. Thus, vortices are found to have fractional statistics parameter $\gamma = \pm \,(2\pi p + \frac{\pi}{m})$, where $p,m \in \mathbb{Z}$. Protocols for probing non-conventional statistics include a mechanical exchange of two anyonic vortices, or time-of-flight measurements \cite{umucalilar2018time,macaluso2019fusion}.\\
 
\paragraph{Flux -- vortex quantisation.}
 We can decompose the order parameter $\psi_{\pm}$ in amplitude and phase
\begin{equation}
\psi_{\pm} = \sqrt{\rho_{\pm}} \, e^{\,i S}
\end{equation}
and use the relation for the current
\begin{equation}
\mathbf{j} = \frac{\hbar}{2M i} \Big[ \psi^{*}_{\pm}(\mathbf{D}\, \psi_{\pm}) - \psi_{\pm} (\mathbf{D}\,\psi_{\pm})^{*}  \Big] = \rho_{\pm} \mathbf{v}_{s} \;,
\end{equation}
to define the superfluid velocity as
\begin{equation}
\mathbf{v}_{s} = \frac{\hbar}{M} \, \Big[ \nabla S - \frac{1}{\hbar}	 \bm{\mathcal{A}} \Big] \;,
\end{equation}
where we recall that $\bm{\mathcal{A}} = \mathbf{A} + \bm{a}\,$. We can now consider the flux generated by $\bm{\mathcal{A}}$ to be $\Phi\,$. Then, if we integrate the circulation around a vortex, we obtain
\begin{equation}\label{eq:circulation}
\omega = \oint_{\mathcal{C}} d\mathbf{r} \cdot \mathbf{v}_{s} = \frac{\hbar}{M} 2\pi n
- \frac{1}{M} \,\Phi \;,
\end{equation}
where $n\in \mathbb{Z}\,$ is the winding number, and $\Phi$ defines a "magnetic" flux. The first term on the r.h.s. is the usual quantisation of circulation for neutral superfluids in units of $h/M\,$ \footnote{Notice this is true only when the integration contour cannot be contracted to a point, \textit{i.e.} when Stokes' theorem is not valid. This requires a vortex profile.}. Now, imposing that at large distances circulation must vanish with $\lim_{\,r\rightarrow \infty} \omega = 0$, then yields
\begin{equation}
0 = \frac{nh}{M} - \frac{\Phi}{M} \;\;\; \Longrightarrow \;\; \Phi = n \,\Phi_{0}  \;,
\end{equation}
where $\Phi_{0} \equiv h$ defines the "magnetic" flux quantum. This is nothing but London's flux quantisation for Abrikosov-Nielsen-Olesen vortices in superconductors. However, it follows from the flux attachment relation \eqref{eq:subeq_1} that not only "magnetic" flux is quantised, but also "charge", meaning
\begin{equation} \label{eq:v_num}
N = \int d^{2} \mathbf{r}\; \rho_{\pm}  \approx \frac{\kappa}{2\pi \hbar}\, \Phi =  \kappa \,n \;,
\end{equation}
which will correspond to a fractional quantisation condition when the Chern-Simons level acts like a filling fraction, in close analogy to the fractional electric charge quantisation found for Laughlin quasiparticles \cite{laughlin1983anomalous}. It is worth noting that in our case the Noether charge corresponds to the number of particles $N$. Alternatively, equation \eqref{eq:v_num} can be regarded as the vortex number over the number of flux quanta attached. Extracting topological charges in ultracold gases is currently possible by means of transport measurements \cite{jotzu2014experimental,aidelsburger2015measuring}, quantised circular dichroism \cite{asteria2019measuring}, Berry curvature reconstruction \cite{flaschner2016experimental}, or variants of quantum state tomography \cite{valdes2019unconventional}. Other recent theoretical proposals involve measurement of the centre-of-mass motion \cite{price2016measurement}. \\

\paragraph{Incompressibility and gapped spectrum.} We note that the interacting terms of the effective action \eqref{eq:final_eff_S} can be rewritten as a conventional Higgs potential of the generic form 
\begin{equation}
V\, (\psi_{\pm}) \sim \Big(1- |\psi_{\pm}|^{2} \Big)^{2} \;.
\end{equation}
We now see that, as it happens in superconductors, there will be an Anderson-Higgs mechanism, and an associated Meissner effect, which is responsible for the incompressibility of the state at certain filling fractions \cite{zhang1995chern, stone1990superfluid}. This would also gap the usual phonon-roton spectrum in superfluids, analogously to the case in superconductors in which the "Higgsed" phonon branch is promoted to the plasma frequency \cite{anderson1963plasmons}. In this case the gapped excitation is a topologically-trivial cyclotron mode, while a magneto-roton branch corresponds to the topological vortices of the theory. 
Once again we refer the reader to Zhang \cite{zhang1995chern} for a thorough discussion and derivation of these and other properties of the ZHK model, such as that of off-diagonal long-range order and Laughlin's wavefunction.

\section{Conclusions} \label{sec:conclusion}

We have investigated whether minimally coupling a gauge potential that is a function of matter density is enough to obtain a gauge theory. By reinterpreting several key aspects of the notion of flux attachment, we found that an Abelian Chern-Simons theory can be expressed in this way. In fact, it is a topological gauge theory. We should note that this already allows us to address several points:

\textit{(i)} Density-dependent gauge potentials are dynamical gauge fields with a non-zero, but trivial, back-action mechanism with matter. Naively, this is not enough to obtain a full gauge theory since its dynamics must be constrained by a local rule, \textit{i.e.} a Gauss's law, restricting the physical states of the system to live in a subregion of the whole Hilbert space at any given point in spacetime. Such constraints are not straightforwardly achieved, and even in that event, the resulting gauge theory could be naively regarded as trivial. This is because its dynamics vanish in the absence of matter, and it is thus not possible to obtain a Maxwell-type theory, which has free propagating modes. 
	
\textit{(ii)} Chern-Simons theory provides an example of a non-trivial gauge theory for which only matter degrees of freedom are needed. This has direct implications for quantum simulation. Typically, two species of atoms are needed for implementing lattice gauge theories, representing matter (sites) and gauge fields (links) respectively. Here we provide an example of a gauge theory in (2+1)D that can be engineered self-consistently with matter only, and with one species. Note that this is also possible for some (1+1)D models by eliminating the gauge fields at the expense of introducing nonlocalities. The peculiarity of Chern-Simons is that the first-order dependence in derivatives, contrary to the usual second-order,  allows for a reduction of the nonlocality even in 2+1 dimensions.
	
\textit{(iii)} The non-triviality of the Chern-Simons theory comes from the intrinsic topological nature, which in the words of Zee \cite{zee2010quantum} "lives in a world without clocks or rulers". We show that a Chern-Simons gauge field is an example of a density-dependent gauge field with a vortex profile. It is then further possible to simulate this using a Bose-Einstein condensate, provided the local constraint and the equation of motion are one and the same. More generally, one might consider a class of density-dependent gauge fields with an arbitrary topological soliton kernel $K^{\,\mu}$, where
\begin{equation}
A^{\,\mu} \, (t, \mathbf{r}) \sim \int d^{d}\mathbf{r'}\; \Big\{ \big[\nabla_{\mathbf{r}\, }K^{\,\mu}\, (\mathbf{r} - \mathbf{r'}) \big]\,\rho\,(t,\mathbf{r'}) \Big\}\;.
\end{equation}

\textit{(iv)} The current view provides some intuition to the apparent conundrum of classifying theories with density-dependent gauge fields. We see that a subclass of them can be related to topological field theories, which fall in between the notion of theories coupled to background gauge potentials, and Yang-Mills type gauge theories. Furthermore, we identify that density-dependent gauge fields naturally appear in some strongly-correlated electron systems like fractional quantum Hall states or gapped quantum spin liquids, and thus, are not only produced synthetically in engineered systems.	\\

We have then proceeded to show how to obtain such a Chern-Simons term at a mean field level, starting from a microscopic weakly interacting system of bosons with internal structure coupled by a light beam. Chern-Simons terms typically emerge at low energies in many-body systems, meaning that they are fictitious, internal, or self-generated. Hence, it is not straight-forward to "derive" such an emergent process. This was possible due to careful design of a Berry connection contribution dependent on the interparticle interactions. This construction was done using a weakly-interacting system contrary to the conventional scenarios in topologically ordered materials, which are strongly coupled. Taking the interactions to be short range, and by constraining the laser configuration, we were able to recover flux attachment. This was then incorporated in the system's action as a constraint, yielding an effective theory for bosonic matter minimally coupled both to a Chern-Simons and a background gauge field. While the origin of the latter contribution is given by conventional artificial gauge fields, the former is singular and density-dependent. Finally, we identified phenomenological consequences of a flux-attached vortex in the Bose fluid, specifically providing a bosonic macroscopic description of fractional quantum Hall states, the ZHK model. 

We emphasise that the relationship with the fractional QHE is a natural consequence of our construction but not the main aim of this work. Both Chern-Simons theory and quantum Hall phenomena have been widely studied in the past. In fact, it has been long known in the context of quantum simulation that it is theoretically possible, although experimentally challenging, to obtain fractional quantum Hall states \cite{sorensen2005fractional,paredes20011,cooper2008rapidly}, either in the lattice or in continuum, by applying a background synthetic gauge field and ramping up interparticle contact interactions, \textit{i.e.} realising an interacting Harper-Hofstadter model. This would emulate the conditions from two-dimensional electron gases where the fractional QHE was originally found, where contact interactions yield the leading order of Coulomb-like interactions. The addition of long-range (\textit{e.g.} dipolar) interactions further stabilises the system \cite{hafezi2007fractional}.  However, the main experimental challenge is the "heating" associated with spontaneous emission, which limits the strength of the applied fields, especially for alkali atoms \cite{spielman2017bose}. While similar challenges possibly also apply to our scheme, several experimental requirements are already available. Optically generated vortices in condensates can be currently induced in multiple ways, for instance via Laguerre-Gauss beams \cite{andersen2006quantized} or using holographic techniques \cite{brachmann2011inducing}. Interatomic interactions can be controlled in ultracold gases by means of Feshbach resonances tuned by magnetic fields \cite{chin2010feshbach}, optically \cite{theis2004tuning}, or by tailoring radio-frequency coupled internal states \cite{sanz2019interaction}. Density-dependent gauge potentials have recently also been realised \cite{clark2018observation,gorg2019realization,schweizer2019floquet,lienhard2020realization}.\\

This view on density-dependent gauge fields is expected to be rather general. Hence, similar ideas to those of this work could be pursued on the lattice \cite{sohal2020intertwined,fradkin1989jordan}, for fermionic systems \cite{lopez1991fractional}, or in other platforms such as helium thin films \cite{chan1974superfluidity} or quantum fluids of light \cite{carusotto2013quantum}. An interesting extension of the current work is based on spinor bosonic condensates \cite{stamper2013spinor, kawaguchi2012spinor}, \textit{e.g.} by considering coherent coupling of three or more internal atomic states. The naive expectation is that the corresponding emergent model would be that of a quantum Hall ferromagnet \cite{ezawa2009quantum}, for which topologically non-trivial spin textures are believed to arise \cite{volovik1988theta,marino2000skyrmion}, namely \textit{baby skyrmions}.

Furthermore, a plethora of new systems could be approached if similar ideas can be extended to a non-Abelian gauge group \cite{goldman2019landau}. Immediate examples of the applicability are the study of non-Abelian FQH states, and the generation of non-Abelian anyonic vortices, for which non-trivial braiding can potentially lead to applications in quantum computing based on topologically protected qubits \cite{nayak2008non}. Additionally, this could also prove useful to theories of gravity in 2+1 dimensions \cite{carlip2005quantum}, since the Einstein-Hilbert action is described by a Chern-Simons theory, indicating that gravity is topological on a planar universe. This can be seen in the so-called first-order formalism by realising that the dreibein and the Lorentz connection act effectively as gauge fields, \textit{i.e.} connections for diffeomorphisms. The realisation and control of such a term coupled to matter would make possible the incorporation of back-action in a consistent way in a quantum simulation of fields in curved spacetime. \\

Our approach to flux attachment explicitly links the presence of interactions at a microscopic level with dynamics of a Berry connection which, in turn, is found to be a Chern-Simons gauge field. It is thus tempting to speculate on whether such a mechanism could take place in real material samples. In that scenario, it would lead to a heuristic picture in which the strength of the interparticle interactions determines the relevance of the Chern-Simons term relative to other scales in the system. The nonlocal interactions caused by the Chern-Simons field could affect the quantum correlations, leading to a correction to the so-called \textit{area law} in the entanglement entropy of the groundstate, which signals the presence of long-range entanglement \cite{chen2010local, zeng2019quantum}, \textit{i.e.}  the topological entanglement entropy is not zero. By extension, \textit{topological order} would arise, even from short-range interactions. \\

\textit{Note Added. --- }  Only recently, we have been aware of the works \cite{correggi2019vortex, lundholm2015average} for which a similar effective theory is considered. We find consistent findings with these studies at the points where both works overlap. Furthermore, the authors provide numerical evidence of the formation of localised density profiles identified as anyons. We also find references \cite{yakaboylu2019quantum, yakaboylu2018anyonic} to be somewhat similar in spirit to our scheme. There, when identical impurities are introduced in a planar bosonic bath, Fr\"ohlich polarons are identified as anyons, which play a similar role to our localised density profiles.

\begin{acknowledgments}
The authors acknowledge helpful discussions with A. Celi, C. Maitland, J. Oll\'e, B. Schroers and L. Tarruell. We also thank K. Hazzard for drawing our attention to Refs. \cite{yakaboylu2019quantum,yakaboylu2018anyonic}. G. V-R. acknowledges financial support from EPSRC CM-CDT Grant No. EP/L015110/1.
\end{acknowledgments}

\appendix

\section{Multivalued Functions}\label{sec:multi}
There is a small subtlety involving the function $\varphi$, which appears when recovering the flux attachment condition $B \propto \rho$ from the expression for the vector potential $\mathbf{A}$. One would naively expect that $\nabla \times \nabla \varphi = 0$, from the conventional vector identities, \textit{i.e.} derivatives commute. However, this is not the case at $r=0$, since $\varphi$ is essentially the polar angle variable, which is multivalued and ill-defined at zero. More generally one would write $\nabla \times \nabla \varphi \,(\mathbf{r})= \alpha \, \delta^{\,(2)}(\mathbf{r})$. To find the proportionality constant $\alpha$ we must integrate on both sides of the last expression for a disk $\mathcal{D}$, of boundary $\partial \mathcal{D}$, and infinitely small radius $\varepsilon \rightarrow 0$, such that
\begin{equation}
\alpha = \int_{\mathcal{D}} d^{2}\mathbf{r} \; \nabla \times \nabla \varphi \,(\mathbf{r}) =  \oint_{\partial\mathcal{D}} dr \; \nabla \varphi = \int^{2\pi}_{0} d\varphi \; \partial_{\,\varphi} \, \varphi = 2\pi \;,
\end{equation}
where we have used the Green-Riemann formula. Hence, we are left with
\begin{equation}
 \epsilon^{\,ij}\,\partial_{\,i} \,\partial_{j}\, \varphi\,(\mathbf{r}) = 2\pi\,\delta^{\,(2)}\, (\mathbf{r}) 
\end{equation} 
 as a final answer. See \cite{kleinert2008multivalued, marino2017quantum} for a more detailed discussion of this result.

\section{Interaction kernel and feasibility }\label{sec:kernel}

When aiming to recover flux attachment from the form of the interaction-induced Berry connection, one might be tempted to identify \eqref{eq:dens_dep_gauge} with expression \eqref{eq:vect_pot_form} for the Chern-Simons vector potential. At least three aspects should be taken into account in following this line of thought: \textit{(i)} The magnetic field $\mathbf{b}_{\pm}$ must be pointing along the $z$ direction, so that it is a "scalar" in the $x-y$ plane. \textit{(ii)} The nonlocal kernel $\mathcal{K}$ requires long-range interparticle interactions. This does not seem a particularly stringent requirement since ultracold dipolar gases present an anisotropic nonlocal interaction kernel $\propto 1/r^{3}$. Yet, we would also require it to yield a $\delta$-function when integrated over the plane, which implies it must be singular. This happens for the Green's function in the Chern-Simons case, for which such kernel is a vortex, allowing
\begin{equation}
\nabla \times \bigg[\frac{\hbar\,\epsilon^{\,ij}}{\kappa}\int d^{2}\mathbf{r'}\; G_{\,j\,}(\mathbf{r} - \mathbf{r'}) \, \rho\,(t,\mathbf{r'}) \bigg] = \frac{2\pi \hbar}{\kappa} \rho\,(t,\mathbf{r})
\end{equation}
to be satisfied. \textit{(iii)} It is worth noting that while the Green's function is a vectorial quantity, the interaction kernel we have considered is scalar. This makes the matching harder than anticipated. For instance, fixing the phase of the laser to be plane-wave like $\phi = \mathbf{k}\cdot \mathbf{r} = k\,(x+y)$, working in Cartesian coordinates, and considering for simplicity $k = k_{x} = k_{y}\,$. We observe that the dynamical contribution to the magnetic field is
\begin{equation} 
\begin{split}
\mathbf{b}_{\pm} =  \pm k \frac{f_{\pm}(\theta) }{8 \Omega} \bigg[ \partial_{x}\int d^{2} \mathbf{r'}\;\mathcal{K}\,(\mathbf{r} - \mathbf{r'})\,\rho_{\pm}(\mathbf{r'}) \\
- \partial_{y}\int d^{2} \mathbf{r'}\;\mathcal{K}\,(\mathbf{r} - \mathbf{r'})\,\rho_{\pm}(\mathbf{r'})\bigg] \hat{z} \;,
\end{split}
\end{equation}
where the magnetic field points in the correct direction. Matching equation \eqref{eq:vect_pot_form} would require the prefactor (dependent on the laser parameters) to be equal to $2\pi \hbar k^{-1}$ and, in addition, the term in square brackets would be set to
\begin{equation}
\begin{split}
\partial_{x}\iint dx'\,dy' \,\frac{x-x'}{(x-x')^{2} + (y-y')^{2}}\,\rho_{\pm}(x',y') \\
- \,\partial_{y}\iint dx'\,dy' \,\frac{-\,(y-y')}{(x-x')^{2} + (y-y')^{2}} \,\rho_{\pm}(x',y')\;,
\end{split}
\end{equation}
constraining the form of the interaction kernel. While solving the constraint would indeed give a magnetic field depending on the matter density which is our end goal. This can be checked numerically. However, at an experimental level it is a significant challenge.

\section{Singular gauge transformation as bosonisation}\label{sec:bosonization}

In this section we closely follow \cite{zhang1995chern}, where a more detailed discussion can be found. Let us consider a microscopic Hamiltonian in 2+1 dimensions of the form 
\begin{equation}
H_{f} = \sum_{j=1}^{N}  \frac{\big| \, \mathbf{p}_{j}-  e \mathbf{A}\,(\mathbf{r}_{j})\,\big|^{2} }{2 m} +\sum_{i<j} V\,(\mathbf{r}_{i} - \mathbf{r}_{j} )+ \sum_{i=1}^{N} e  A_{0}\,(\mathbf{r}_{i}) \; ,
\end{equation}
which involves minimal coupling to a gauge field and a pairwise interaction potential $V$. Hamiltonian $H_{f}$ satisfies the time-independent Schr\"odinger equation 
\begin{equation}\label{eq:fermion_schro}
H_{f} \, \Psi \, (\mathbf{r}_{1}, \dots , \mathbf{r}_{N}) = E\, \Psi \,(\mathbf{r}_{1}, \dots , \mathbf{r}_{N})\;,
\end{equation}
where $\Psi$ is a totally antisymmetric many-body wavefunction. Thus, this is a fermionic problem. In a similar spirit, we can define a new Hamiltonian
\begin{equation}
H_{b} = \sum_{j=1}^{N}  \frac{\big| \, \mathbf{p}_{j}-  e \mathbf{A}\,(\mathbf{r}_{j}) -  e \bm{a}\,(\mathbf{r}_{j})\,\big|^{2} }{2 m} +\sum_{i<j} V\,(\mathbf{r}_{i} - \mathbf{r}_{j} )+ \sum_{i=1}^{N} e A_{0}\,(\mathbf{r}_{i}) \; ,
\end{equation} 
where $\bm{a}$ is a vector field yet to be defined. Hamiltonian $H_{b}$ satisfies the eigenvalue equation 
\begin{equation}\label{eq:boson_schro}
H_{b} \, \Phi \,(\mathbf{r}_{1}, \dots , \mathbf{r}_{N}) = E'\, \Phi \,(\mathbf{r}_{1}, \dots , \mathbf{r}_{N})\;,
\end{equation}
where now $\Phi$ is a totally symmetric wavefunction, so the problem is bosonic in nature. The claim is that, while one would naively think that equations \eqref{eq:fermion_schro} and \eqref{eq:boson_schro} describe completely unrelated problems, there exist a canonical transformation that maps one into the other. Consider the relation 
\begin{equation}
\tilde{\Psi} \, (\mathbf{r}_{1}, \dots , \mathbf{r}_{N}) = \Big[ e^{-i \frac{\hbar}{\kappa} \sum_{i<j} \alpha\,(\mathbf{r}_{i} - \mathbf{r}_{j})}  \Big] \, \Psi \, (\mathbf{r}_{1}, \dots , \mathbf{r}_{N}) \, ,
\end{equation}
where $\alpha$ defines the angle formed by the direction $|\mathbf{r}_{i} - \mathbf{r}_{j}|$ between two particles in the system, and an arbitrary reference direction. The term in square brackets is a unitary matrix $U$ and can be alternatively represented in complex coordinate notation as 
\begin{equation}
U = - \frac{\hbar}{\kappa} \prod_{i<j} \frac{z_{i} - z_{j}}{|z_{i} - z_{j}|} \;\;\;\;\;\;\;\;\mathrm{for}\;\;\;\;\;\;\;\; z = x + i\, y \;.
\end{equation}
This is a \textit{singular} gauge transformation, analogous to that of the Aharonov-Bohm bound state problem, where $\mathbf{A}\,(\mathbf{r}) \sim \nabla \, \mathrm{arg} (\mathbf{r})$ is a pure gauge vector potential which can be removed by means of a gauge transformation $\psi' \sim \exp{[\,i \,\mathrm{arg} (\mathbf{r}) \,]}  \, \psi$. Notice that there is an implicit hardcore constraint in the transformation involving $U$ since it is ill-defined at $\mathbf{r}_{i}=\mathbf{r}_{j}$. Let us transform the fermionic Hamiltonian
\begin{equation}
\tilde{H}_{f} =  U \, H_{f} \, U^{-1} \; ,
\end{equation}
where the key term consists of 
\begin{equation}
\mathbf{p}_{j} -  e \mathbf{A}\,(\mathbf{r}_{i}) -  e \bm{a}\,(\mathbf{r}_{i}) =  U \, \big[  \mathbf{p}_{i}-  e \mathbf{A}\,(\mathbf{r}_{i}) \big] \, U^{-1} \; .
\end{equation}
Here, $\bm{a}$ constitutes a many-body version of the Aharonov-Bohm vector potential, defined as
\begin{equation}
 e \bm{a}\,(\mathbf{r}_{i}) \equiv \frac{\hbar}{\kappa} \sum_{j \neq i} \nabla_{\mathbf{r}_{i}} \alpha\,(\mathbf{r}_{i} - \mathbf{r}_{j}) = \frac{\hbar}{\kappa} \sum_{j \neq i} \hat{z} \times \frac{\mathbf{r}_{i} - \mathbf{r}_{j}}{|\mathbf{r}_{i} - \mathbf{r}_{j}|^{2}}  \, .
\end{equation}
Now, $\tilde{H}_{f}$ has exactly the same form as $H_{b}$, but it defines a different eigenvalue problem unless $\tilde{\Psi} = \Phi\,$. That is, unless the statistics of the originally antisymmetric wavefunction $\Psi$ become symmetric after the canonical transformation. Provided the property $\alpha\,(\mathbf{r}_{i} - \mathbf{r}_{j}) = \pi + \alpha\,(\mathbf{r}_{j} - \mathbf{r}_{i})$ is fulfilled, it can be verified that upon exchange of two particles at different positions
\begin{equation}
\begin{split}
\tilde{\Psi} \,(\mathbf{r}_{1}, \dots , \mathbf{r}_{i} , \dots, \mathbf{r}_{j} , \dots, \mathbf{r}_{N})\\ = -\, e^{\,i \frac{\pi}{\kappa}} \, \tilde{\Psi} \,(\mathbf{r}_{1}, \dots , \mathbf{r}_{j} , \dots, \mathbf{r}_{i} , \dots, \mathbf{r}_{N}) \,
\end{split}
\end{equation}
the many-body wavefunction acquires a phase factor in addition to the usual fermionic sign. This new contribution indeed comes from the Aharonov-Bohm effect. We observe that for values $\kappa = 1/(2m +1)$ where $m \in \mathbb{Z}$, the transformed wavefunction becomes bosonic, meaning that $\tilde{\Psi} =\Phi$, and therefore equations \eqref{eq:fermion_schro} and \eqref{eq:boson_schro} describe the same eigenvalue problem. For $\kappa = 1/(2m) $ the system is fermionic and, for any other value, it is regarded as anyonic. The presence (absence) of the vector potential $\bm{a}$ is induced (removed) by the singular gauge transformation performed by $U$ at the expense of effectively changing the statistics of the problem. Thus, this process describes an operator bosonisation or fermionisation mechanism. The connection to flux attachment is made by taking the curl over such a vector potential to verify
\begin{equation}
b\,(\mathbf{r}_{i}) = \frac{2 \pi \hbar}{\kappa e} \sum_{j \neq i} \delta^{\,(2)} (\mathbf{r}_{i} - \mathbf{r}_{j} ) \equiv \frac{2 \pi \hbar}{\kappa e^{2}} \rho \,(\mathbf{r}_{i}) \, .
\end{equation}

\bibliographystyle{apsrev4-2}
\bibliography{refer, CS_VorticesNotes}

\end{document}